\documentclass[aps,prl,twocolumn,showpacs,a4paper,superscriptaddress]{revtex4-1}
\usepackage{graphicx,bm,amsmath,dcolumn,amssymb}
\usepackage{longtable}
\graphicspath{{figures}}
\usepackage{geometry,hyperref,color}
\begin{document}
\title{Dynamical Quantum Phase Transition from Critical Quantum Quench}
%\title{Critical Quantum Quench}
  \author{Chengxiang Ding}
  \email{dingcx@ahut.edu.cn}
  \affiliation{School of Science and Engineering of Mathematics and Physics, Anhui University of Technology, Maanshan 243002, China }
\date{\today}
\begin{abstract}

  We study the dynamical quantum phase transition of the critical quantum quench, 
  in which the prequenched Hamiltonian, or the postquenched Hamiltonian,
  or both of them are set to be the critical points of equilibrium quantum phase transitions,
  we find half-quantized or unquantized dynamical topological order parameter and dynamical Chern number;
  these results and also the existence of dynamical quantum phase transition are all closely related to the singularity 
  of the Bogoliubov angle at the gap-closing momentum. 
  The effects of the singularity may also be canceled out if both the prequenched and postquenched Hamiltonians are critical, 
  then the dynamical topological order parameter and dynamical Chern number restore to integer ones.
  Our findings show that the widely accepted definitions of dynamical topological order parameter and dynamical Chern number are
  problematic for the critical quenches in the perspective of topology, which call for new definitions of them.
 
\end{abstract}
%\pacs{05.50.+q, 64.60.Cn, 64.60.Fr, 75.10.Hk}
\maketitle 
{\it Introduction.}---The theory of dynamical quantum phase transition\cite{Heyl2013} (DQPT) concerns the dynamical behaviors, 
especially the nonanalytic behaviors, of a many-body system after a sudden change of the parameters (quantum quench).
Generally, the phase-transition-like nonanalytic behaviors of the so called dynamical free energy is 
at a series of critical times when the overlap between the initial state and the evolving state is zero. 
The conception of scaling and universality, and the related renormalization theory, are also applicable for the DQPT\cite{univ,univ1,univ2,univ3}.
A lot of DQPTs exhibit linear singularity with critical exponent $\alpha=1$, however, nonlinear singularity has also 
been found recently\cite{univ3}. 
DQPT has been studied in different systems\cite{extIsing, XYDQPT,kitaev,rand, creutz} and been generalized to 
mixed states\cite{mixed1,mixed2}, open system\cite{open}, Floquet system\cite{floquet}, 
and so forth\cite{sw,ulink,sc,excited,nonint}.
It can also be realized in in the experiments of ultra-cold-atomic gases\cite{cold1, cold2} and trapped ions\cite{trap1,trap2,trap3}.

Different from the symmetry breaking phase transition in equilibrium system, the DQPT has no local order parameter, 
however, it can be characterized by a dynamical topological order parameter\cite{DTOP} (DTOP), which is extracted from 
the  Pancharatnam phase of the Loschomidt amplitude. Generally, the DTOP is quantized as integer number and changes its value at the critical times, 
it is applicable for describing the topology of a quantum quench at a given time. 
In addition, there is another type of topological invariant for quantum quench,
which is defined in the ($\vec{k}, t$) space by mapping it to the Bloch sphere\cite{DPchern}; in a two-dimensional 
Chern insulator, this is a Hopf invariant. Such type of topological invariant is also generalized to
one-dimensional system, and a dynamical Chern number is defined\cite{DPchern1}.

In this paper, we consider the DQPT of a special case of quantum quench, in which the prequenched Hamiltonian, or the postquenched Hamiltonian,
or both of them are set to be the critical points of equilibrium quantum phase transitions, paying special attention to the 
topological properties. We find that the DTOP and dynamical Chern number can be half-quantized or unquantized,
we demonstrate that these results and also the existence of DQPT are closely related to the singularity of the Bogoliubov angle of the Bloch vector
at the gap-closing momentum.

We consider a XY chain in a transverse magnetic field
\begin{eqnarray}
H=\frac{1}{2}\sum\limits_{j=1}^N\Big[\frac{1+\gamma}{2}\sigma_j^x\sigma_{j+1}^x+\frac{1-\gamma}{2}\sigma_j^y\sigma_{j+1}^y-g\sigma_j^z\Big].
\end{eqnarray}
By the Jordan-Wigner transformation and Fourier transformation, the model can be transformed to 
\begin{eqnarray}
H(\gamma,g)=\sum\limits_{k>0} \eta_k^\dag h(k) \eta_k \label{hk}
\end{eqnarray}
where $\eta_k=(c_k, c_{-k}^\dag)$ and $h(k)=\vec{d}(k)\cdot\vec{\sigma}$, with $\vec{\sigma}$ the Pauli matrix
and $\vec{d}=(0,\gamma\sin k, \cos k-g)$ the Bloch vector.
Diagonalization of $h(k)$ yields the dispersion relation $\epsilon_k(\gamma,g)=\sqrt{(\cos k-g)^2+\gamma^2\sin^2k}$.

Prepare an initial state $|\psi_0\rangle$, which is a ground state of a prequenched Hamiltonian $H_i=H(\gamma_i,g_i)$, and then suddenly 
change the system to a postquenched Hamiltonian $H_f=H(\gamma_f,g_f)$, the system may undergo a DQPT, whose singularity is reflected 
in the rate function $l(t)=-\lim_{N\rightarrow\infty}\frac{1}{N}\log|\mathcal{L}(t)|^2$ 
and the dynamical free energy $f(z)=-\lim_{N\rightarrow\infty}\frac{1}{N}\log\mathcal{Z}(z)$
at a series of critical times.
Here, the boundary function $\mathcal{Z}(z)=\langle\psi_0|e^{-zH_f}|\psi_0\rangle$
is an analytic continuation of the Loschomidt amplitude $\mathcal{L}(t)=\langle\psi_0|e^{-itH_f}|\psi_0\rangle$ under $z={\rm Re}z+it$,
and the definition of $\mathcal{L}(t)$ is similar to fidelity\cite{fidelity}.
For the XY model, $\mathcal{Z}(z)$ can be calculated analytically, which is $\mathcal{Z}(z)=\prod_{k>0}\mathcal{Z}_k(z)$, with 
\begin{eqnarray}
\mathcal{Z}_k(z)=\cos^2\varphi_k e^{\epsilon_k(\gamma_f, g_f)z}+\sin^2\varphi_k e^{-\epsilon_k(\gamma_f,g_f)z}, \label{Zk}
\end{eqnarray}
where $\varphi_k=\theta_k(\gamma_i,g_i)-\theta_k(\gamma_f,g_f)$, with $\tan[2\theta_k(\gamma,g)]\overset{\rm def}=\gamma\sin k/(g-\cos k)$, 
$\theta_k\in[0,\pi/2]$.
The Fisher zeros of $f(z)$ are $z_n=1/[2\epsilon_k(\gamma_f,g_f)]\cdot[\ln\tan^2\varphi_k+i\pi(2n+1)]$, with $n=0,1,2,\cdots$;
when $\varphi_k=\pm\pi/4$, namely the initial Bloch vector $\vec{d}_i$ and the final Bloch vector $\vec{d}_f$ are 
perpendicular to each other, the real parts of $z_n$ are zero, and we get the critical times $t_n=t^*(n+1/2)$,
with $t^*=\pi/\epsilon_{k^*}(\gamma_f,g_f)$,
% i.e, $\vec{d}_i\cdot\vec{d}_f=0$, 
where $k^*$ is determined by 
\begin{eqnarray}
\vec{d}_i\cdot\vec{d}_f=(\cos k^*-g_i)(\cos k^*-g_f)+\gamma_i\gamma_f\sin^2 k^*=0. \label{kstar}
\end{eqnarray}

\begin{figure}[htpb]
\centering
\includegraphics[width=0.9\columnwidth]{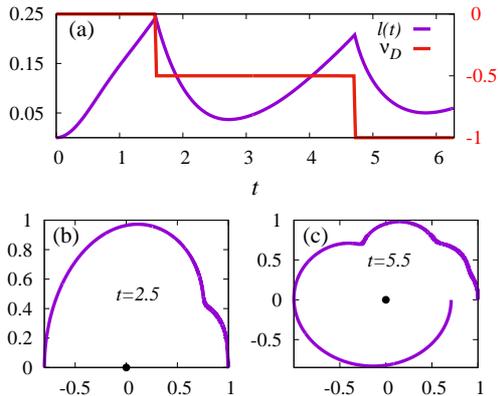}
\caption{(Color online) (a) DQPT from $(\gamma_i,g_i)=(1,1)$ to $(\gamma_f,g_f)=(1,2)$; 
(b),(c) trajectories of the vector $\vec{r}_k$ for this DQPT at $t=2.5$ and $t=5.5$, respectively.}
\label{figA}
\end{figure}

{\it Quench from critical point to noncritical point}.---If the prequenched Hamiltonian $H_i$ 
is a critical point of an equilibrium quantum phase transition but the postquenched Hamiltonian $H_f$ is not, 
there may be a DQPT, as shown in Fig. \ref{figA}(a).
We can see that the DTOP $\nu_D(t)$ is always the integer times of a half-quantized value 1/2.
%this is much different from the noncritical quench; 
%it is very similar to the fractonal winding number of Kitaev chain with long-range paring \cite{lkitaev}.

DTOP\cite{DTOP} is defined in terms of the Pancharatnam geometrical phase of the Loschomidt amplitude
\begin{eqnarray}
\nu_D(t)=\frac{1}{2\pi}\oint_0^\pi\frac{\partial \phi^G_k(t)}{\partial k}dk, \label{vD}
\end{eqnarray}
where $\phi^G_k(t)=\phi_k^\mathcal{L}(t)-\phi^{\rm dyn}_k(t)$, with $\phi_k^{\mathcal{L}}(t)$
the phase of the Loschomidt amplitude at momentum $k$ and $\phi^{\rm dyn}_k(t)$ the dynamical phase
\begin{eqnarray}
\mathcal{L}_k(t)&=&\mathcal{Z}_k(it)=|\mathcal{L}_k(t)|e^{i\phi_k^\mathcal{L}(t)}, \label{Lk}\\
\phi^{\rm dyn}_k(t)&=&-\int_0^t ds\langle\psi_k(s)|h_f(k)|\psi_k(s)\rangle\nonumber\\
                &=&\epsilon_k(\gamma_f,g_f)t\cos(2\varphi_k).\label{dyn}
\end{eqnarray}
It is obvious that $\nu_D$ can be understood as the winding number of the vector 
\begin{eqnarray}
\vec{r}_k=(x_k, y_k)=|\mathcal{L}_k(t)|e^{i\phi^G_k(t)} \label{rk}
\end{eqnarray}
around the origin.
For the noncritical quench, the trajectory of vector $\vec{r}_k$ is a closed loop as $k$ varies from 0 to $\pi$,
the reason lies in the fact that $\vec{r}_{k\rightarrow 0}$=$\vec{r}_{k\rightarrow \pi}$, which defines an effective Brillouin zone $[0, \pi]$.
However, for the critical quench, the trajectory in general is not closed, becase $\vec{r}_{k\rightarrow 0}$
can be unequal to $\vec{r}_{k\rightarrow\pi}$ for the critical quantum quench. This will make the definition of Eq. (\ref{vD}) problematic 
in the perspective of topology, which will be discussed in detail later.
Note that $k\rightarrow0$ and $k\rightarrow\pi$ are the two momenta that make $\phi_k^G(t)$ always be $n\pi$,
we call the two momenta `fixed points', which are also the fixed points of the evolving Bloch vector defined in Eq. (\ref{dt}).
If $\vec{r}_{k\rightarrow 0}\ne\vec{r}_{k\rightarrow\pi}$, the two corresponding points of $\vec{r}_k$
will be two different positions on the real axis, so the trajectory is not closed. 
The difference of $\vec{r}_{k\rightarrow 0}$ and $\vec{r}_{k\rightarrow\pi}$ stems from the 
singularity of the Bogoliubov angle at the gap-closing momentum $k_c$.
Take the quench shown in Fig. \ref{figA}(a) as an example,
the gap-closing momentum of $H_i$ is $k_c=0$, where $\theta_{k\rightarrow0}(\gamma_i,g_i)=\pi/4$ 
and $\theta_{k\rightarrow 0}(\gamma_f,g_f)=0$, so $\varphi_{k\rightarrow0}=\pi/4$,
this leads to $\vec{r}_{k\rightarrow0}=\cos[\epsilon_0(\gamma_f,g_f)t]$ according to Eqs. \ref{Zk}, \ref{Lk}, \ref{dyn}, and \ref{rk};
however similar analysis for $k\rightarrow\pi$ gives $\vec{r}_{k\rightarrow\pi}=1$, 
so $\vec{r}_{k\rightarrow 0}$ is not equal to $\vec{r}_{k\rightarrow\pi}$ except at some special time $t=2n\pi/\epsilon_0(\gamma_f,g_f)=2n\pi$.
Two examples are shown in Fig. \ref{figA}(b) and (c);
in Fig. \ref{figA}(b), the angle swept by the vector $\vec{r}_k$ is $-\pi$, thus it gives a winding number of $\nu_D=-1/2$;
however, in Fig. \ref{figA} (c), the angle swept by $\vec{r}_k$ is $-2\pi$, so the winding number is $\nu_D=-1$;
in both cases, the trajectories of $\vec{r}_k$ are not closed. 
It is obvious that the angle swept by $\vec{r}_k$ is related to the position of the origin, in Fig. \ref{figA}(b),
it is between the two points of $\vec{r}_{k\rightarrow 0}$ and $\vec{r}_{k\rightarrow \pi}$, therefore there is a discontinuous change 
of the Pancharatnam geometrical phase $\phi^G_k$ at the fixed points, this is the core origination of the half quantization of DTOP.

\begin{figure}[htpb]
\centering
\includegraphics[width=1.0\columnwidth]{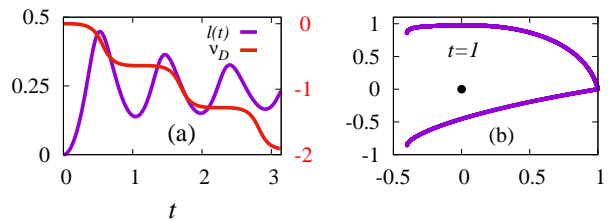}
\caption{(Color online) (a) quantum quench from $(\gamma_i,g_i)=(0,0.5)$ to $(\gamma_f,g_f)=(3,1.5)$, 
note that the peaks of $l(t)$ are not in one-to-one correspondence to the inflexions of the DTOP;
(b) trajectory of the vector $\vec{r}_k$ for this quench at $t=1$.}
\label{figB}
\end{figure}

Now let us consider another example, in which the prequenched Hamiltonian is at the XX chain, i.e. $\gamma_i=0$; 
in this case, there is no DQPT, and the DTOP is unquantized, as shown in Fig. \ref{figB}(a).
The absence of DQPT is owing to the singularity of the Bogoliubov angle at the gap-closing momentum $k_c=\pi/3$.
Here $k_c$ happens to be the same as the momentum $k^*$ that satisfies Eq. (\ref{kstar}), 
however, this does not mean there will be a DQPT, because at this point, the Bogoliubov angle is ill defined, 
so we should take the limit $k\rightarrow k^*$, which gives $\varphi_{k\rightarrow k^*_{-}}\approx0.3085\pi$ 
and $\varphi_{k\rightarrow k^*_{+}}\approx-0.1915\pi$, both of them are not equal to $\pm\pi/4$. 
In fact, in the whole effective Brillouin zone $[0,\pi]$, we can not find a momentum $k^*$ that satisfies $\phi_{k\rightarrow k^*}=\pm\pi/4$,
so the Fisher zeros never intersect the imaginary axis, and the rate function always has no singularity.
Here we can see the particularity of the critical quench, the existence of a $k^*$ that satisfies Eq. (\ref{kstar}) does not mean
$\varphi_{k^*}=\pm\pi/4$, i.e., the existence of a DQPT; 
here Eq. (\ref{kstar}) is satisfied only because of the fact that $\vec{d}_i$ is zero at the gap-closing point.
In contrary, in the noncritical quench, 
the condition of Eq. (\ref{kstar}) and $\varphi_{k^*}=\pm\pi/4$ are the same meaning.

Also because of the singularity of the Bogoliubov angle, the trajectory of vector $\vec{r}_k$ is unclosed, as shown in Fig. \ref{figB}(b). 
%$\theta_{k\rightarrow\frac{\pi}{3}_+}(\gamma_i,g_i)=0$ but $\theta_{k\rightarrow\frac{\pi}{3}_-}(\gamma_i,g_i)=\pi/2$, 
$\theta_{k\rightarrow(\pi/3)_+}(\gamma_i,g_i)=0$ but $\theta_{k\rightarrow(\pi/3)_-}(\gamma_i,g_i)=\pi/2$, 
this will lead to the discontinuous change of $\phi_k^G$ and consequently 
the discontinuous change of $\vec{r}_k$ at $k=\pi/3$, so the trajectory is not closed, i.e., $\nu_D$ is not quantized.

 Up to now, the DTOP has been calculated according to Eq. (\ref{vD}), 
 it is well defined only when the effective Brillouin zone $[0, \pi]$ is endowed with a unit circle topology.
 However, for the critical quench studied in this paper, this condition is not satisfied due to the singularity of the Bogoliubov angle 
 (i.e., the singularity of the critical state), and the integral of Eq. (\ref{vD}) is not over a closed manifold.
 This problem is a kind of topological singularity, which may have different sources.
 A very similar problem is the fractional topological number in the long-range interaction Kitaev chain\cite{lkitaev};
 in that case, the trajectory of the winding vector is also unclosed.
 A simple way to deal with the situation in the long-range Kitaev chain is to define 
 a new winding vector with its angle twice that of the original one.
 For the case in Fig. \ref{figA} of current paper, since the length of the winding vector $\vec{r}_k$ is not equal 
 in the cases of $k\rightarrow0$ and $k\rightarrow\pi$, more processing is needed to make the lengths of the two cases equal, 
 such as 
 \begin{eqnarray}
 \vec{R}_k=R_ke^{2i\phi^G_k(t)}. \label{Rk}
 \end{eqnarray}
 Here $R_k$ is the length of $\vec{R}_k$, which is mapped from $r_k$, it satisfies $R_{k\rightarrow0}=R_{k\rightarrow\pi}$.
 For example, this mapping can be $R_k=r_k+(r_{k\rightarrow\pi}-r_{k\rightarrow0})\cdot(\pi-k)/\pi$. 
 Then the DTOP can still be defined on a closed manifold
 \begin{eqnarray}
 \nu_D(t)=\frac{1}{4\pi} \oint_0^\pi \frac{R_x\partial_k R_y-R_y\partial_k R_x}{R_k^2} dk , \label{vD1}
 \end{eqnarray}
 where $R_x$ and $R_y$ are the two components of $\vec{R}_k$, i.e., $\vec{R}_k=(R_x,R_y)$.
 Figures \ref{Rkfig}(a) and (b) show the trajectories of $\vec{R}_k$ in correspondence to the cases of Figs. \ref{figA}(b) and (c).
 It should be noted that the integral is divided by $2\pi$ in Eq. (\ref{vD}) but $4\pi$ in Eq. ({\ref{vD1}), 
 the reason is that the angle of $\vec{R}_k$ is doubled from the angle of $\vec{r}_k$, and this multiple should be removed.
 For the case of Fig. \ref{figB}, the idea is similar, 
 but the multiple is no longer an integer, it also should be removed, and the final conclusion remains unchanged, it is still unquantized.
 \begin{figure}[htpb]
 \centering
 \includegraphics[width=1.0\columnwidth]{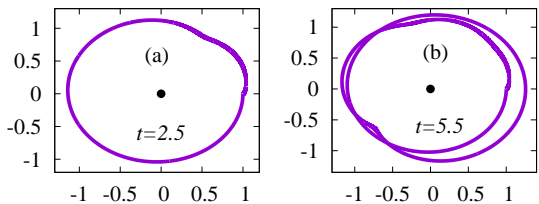}
 \caption{(Color online) Trajectories of $\vec{R}_k$ at $t=2.5$ and $t=5.5$ for the DQPT from $(\gamma_i,g_i)=(1,1)$ to $(\gamma_f,g_f)=(1,2)$.}
 \label{Rkfig}
 \end{figure}

The singularity of the Bogoliubov angle can also lead to the half-quantized and unquantized dynamical Chern number.
For a quench from $h_i(k)$ to $h_f(k)$, where $h(k)$ is defined in Eq. (\ref{hk}), 
no matter whether there is a DQPT, a dynamical Chern number\cite{DPchern1} can be defined as 
\begin{eqnarray}
C^m_{\rm dyn}=\frac{1}{4\pi}\int_{k_m}^{k_{m+1}}dk\int_0^{\frac{\pi}{\epsilon_k(\gamma_f,g_f)}}dt(\hat{d}^t\times\partial_t\hat{d}^t)\cdot\partial_k \hat{d}^t, \label{chern}
\end{eqnarray}
where $\hat{d}^t$ is the Bloch vector of the evolving density matrix $\rho_t=e^{-ih_f(k)t}\rho_i e^{ih_f(k)t}=\frac{1}{2}[1-\hat{d}^t\cdot\vec{\sigma}]$,
and $\rho_i=|\psi_0(k)\rangle\langle\psi_0(k)|=\frac{1}{2}[1-\hat{d}_i\cdot\vec{\sigma}]$ is the initial density matrix,
with $|\psi_0(k)\rangle$ the ground state of the initial Hamiltonian $h_i(k)$, and $\hat{d}_i=\vec{d}_i/|\vec{d}_i|$.
The evolving Bloch vector $\vec{d}^t$ can be calculated analytically\cite{DPchern1},
\begin{eqnarray}
 \hat{d}^t=&&\hat{d}_i\cos(2|\vec{d}_f|t)+2\hat{d}_f(\hat{d}_i\cdot\hat{d}_f)\sin^2(|\vec{d}_f|t) \nonumber\\
&&-\hat{d}_i\times\hat{d}_f\sin(2|\vec{d}_f|t). \label{dt}
\end{eqnarray}
In Eq. (\ref{chern}), $k_m$ is the $m$-th fixed point of $\vec{d}^t$, where the Bloch vector $\vec{d}^t$ keep still during the time evolution.
Generally, at such fixed point, for a noncritical quantum quench,
the initial Bloch vector $\vec{d}_i$ and $\vec{d}_f$ are parallel (or antiparallel) to each other.

For the critical quench shown in Fig. \ref{figA}, the fixed points are $k_1=0$ and $k_2=\pi$.  For the second fixed point $k=\pi$, 
the initial Bloch vector $\vec{d}_i$ and the final Bloch vector $\vec{d}_f$ are parallel to each other,
i.e., $\varphi_{k=\pi}=0$, this is similar to the noncritical quench.
However, the first fixed point $k=0$ is very special, it is a gap-closing point,
in such point we should take the limit $k\rightarrow 0_+$ instead of setting $k=0$, 
in this limit the Bogoliubov angle of $\vec{d}_i$ is $\theta_{k\rightarrow 0_+}(\gamma_i,g_i)=\pi/4$ and the 
angle of $\vec{d}_f$ is $\theta_{k\rightarrow 0_+}(\gamma_f,g_f)=0$, therefore $\varphi_{k\rightarrow 0_+}=\pi/4$.
%thus the two Bloch vectors are not parallel but perpendicular to each other;
%the evolving vector $\hat{d}^t$ can keep still during the time evolution 
%is owing to the fact that at such gap-closing momentum $\vec{d}_i$ is zero. 
In fact, we can find that the range of $\varphi_k$ is $[0,\pi/4]$ as $k$ varies from $k_1$ to $k_2$, 
which is only half of that of the noncritical quench, accordingly, the vector $\hat{d}^t$ sweeps only half of the Bloch sphere,
as shown in Fig. \ref{half_chern}, the corresponding solid angle is only $2\pi$, therefore the dynamical Chern number is 1/2.
\begin{figure}[htpb]
\centering
\includegraphics[width=0.5\columnwidth]{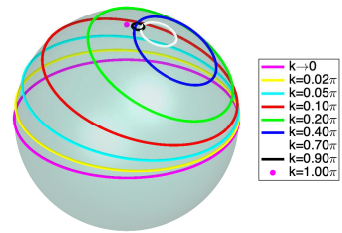}
\caption{(Color online) Time evolution of the Bloch vector $\hat{d}^t$ of different momenta for 
the quantum quench from $(\gamma_i,g_i)=(1,1)$ to $(\gamma_f,g_f)=(1,2)$. }
\label{half_chern}
\end{figure}

For the quench shown in Fig. \ref{figB}, the gap-closing momentum is $k_c=\pi/3$, this is a singular point with 
$\theta_{k\rightarrow\frac{\pi}{3}_-}(\gamma_i,g_i)=\pi/2\ne\theta_{k\rightarrow\frac{\pi}{3}_+}(\gamma_i,g_i)=0$, this not only leads to the 
unquantized DTOP but also the unquantized dynamical Chern number. As shown in Fig. \ref{unquan}, the trajectories of $k\rightarrow\frac{\pi}{3}_-$ and 
$k\rightarrow\frac{\pi}{3}_+$ divide the surface of the Bloch sphere into three parts, the area between the two trajectories are never swept
by the Bloch vector $\hat{d}^t$. The solid angle in correspondence to the swept area is $C_{\rm dyn}\cdot4\pi$, with $C_{\rm dyn}\approx0.64$
obtained from Eq. (\ref{chern}).
\begin{figure}[htpb]
\centering
\includegraphics[width=0.5\columnwidth]{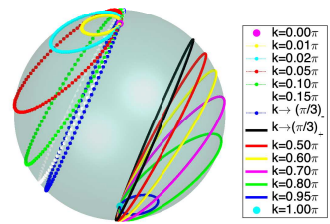}
\caption{(Color online) Time evolution of the Bloch vector $\hat{d}^t$ of different momenta for 
the quantum quench from $(\gamma_i,g_i)=(0,0.5)$ to $(\gamma_f,g_f)=(3,1.5)$.}
\label{unquan}
\end{figure}

From the two examples, we can see that the discontinuous change of the Bogoliubov angle at the gap-closing momentum makes the mapping from the 
space $(k,t)$ to the surface of the Bloch sphere incomplete, so the dynamical Chern number is not integer.

Unquantization of DTOP and dynamical Chern number does not always correspond to the absence of DQPT, 
for example, in the quench from $(\gamma_i,g_i)=(0,0.5)$ to $(\gamma_f,g_f)=(0.5,0.8)$, 
although the DTOP and the dynamical Chern number are not quantized, there still exists a DQPT, 
and the unquantized DTOP still can be used as a detector of the DQPT, as shown in Fig. \ref{cncc}(a).
In fact, in this example, there are two $k^*$ satisfy Eq. (\ref{kstar}), one is $k_1^*=\pi/3$, which coincides with the gap-closing momentum $k_c$,
another one is $k_2^*=\arccos0.8$. From the examples shown in Fig. \ref{figB} and \ref{unquan}, 
we know that the coincidence of $k_1^*$ and $k_c$ can lead to 
the absence of singularity in the rate function $l(t)$ and the unquantization of DTOP and dynamical Chern number, 
however, here $k_2^*$ can recover the singularity in $l(t)$; the final result is the interplay of the effects of $k_1^*$ and $k_2^*$, 
so we get the results shown in Fig. \ref{cncc}(a).
\begin{figure}[htpb]
\centering
\includegraphics[width=1.0\columnwidth]{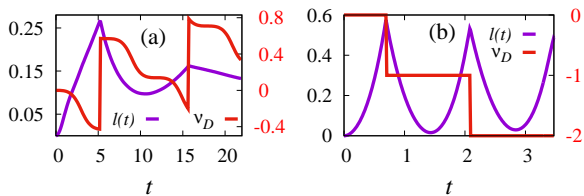}
\caption{(Color online) (a) DQPT from $(\gamma_i,g_i)=(0,0.5)$ to $(\gamma_f,g_f)=(0.5,0.8)$;
 (b)  DQPT from $(\gamma_i,g_i)=(2,1)$ to $(\gamma_f,g_f)=(-2,1)$.}
\label{cncc}
\end{figure}

{\it Quench from noncritical point to critical point}.---If the prequenched Hamiltonian is noncritical but the postquenched Hamiltonian is critical,
there is no DQPT, %a typical example is shown in fig. ?.  
such as the quench from $(\gamma_i,g_i)=(1,0.5)$ to $(\gamma_f,g_f)=(1,1)$.
In this case the momentum that satisfies Eq. (\ref{kstar}) is $k=0$, 
which is also the gap-closing point of the postquenched Hamiltonian, i.e., $\epsilon_0(\gamma_f,g_f)=0$, so the critical time $t^*=\infty$,
 therefore we can not find a DQPT in finite time.
However in this case, we can still find a half-quantized dynamical Chern number, similar to the case shown in Fig. \ref{half_chern}.

{\it Quench from critical point to critical point }.---If both the prequenched and postquenched Hamiltonian are critical, 
the situation becomes a little subtle. 
If the gap-closing momentum of the prequenched Hamiltonian is different from that of the postquenched Hamiltonian, we may get a DQPT with
half-quantized DTOP, such as the quench from $(\gamma_i,g_i)=(1,1)$ to $(\gamma_f,g_f)=(0,0.5)$,
which is similar to the case shown in Fig. \ref{figA}.
However, if the gap-closing momentum of the prequenched Hamiltonian and the 
postquenched Hamiltonian are the same, then the singularity of the two Bogoliubov angles can cancel out with each other,
and the DTOP and dynamical Chern number restore to integer ones, 
such as the quench from $(\gamma_i,g_i)=(2,1)$ to $(\gamma_f,g_f)=(-2,1)$, shown in Fig. \ref{cncc}(b). 
In this case, there are two $k^*$ satisfy Eq. (\ref{kstar}), which are $k_1^*=0$ and $k_2^*=\arccos(-3/5)$; $k_1^*$ coincides
with the gap-closing points of both the prequenched Hamiltonian and postquenched Hamiltonian, 
it does not lead to any singularity in the rate function $l(t)$, however $k_2^*$ can lead to certain singularity
in $l(t)$, therefore there is a DQPT. For the DTOP, because both $\epsilon_0(\gamma_i,g_i)$ and $\epsilon_0(\gamma_f,g_f)$
are zero, we know $\vec{r}_{k\rightarrow 0}=1$ from Eqs. (\ref{Zk}) and (\ref{rk}), which is the equal to $\vec{r}_{k\rightarrow \pi}$, 
namely the values of $\vec{r}_k$ are the same at the two fixed points, thus the trajectory of $\vec{r}_k$ is always closed, 
i.e., $\nu_D$ is an integer. For the dynamical Chern number, it is easy to find that the range of the angle difference between 
$\theta_k(\gamma_i,g_i)$ and $\theta_k(\gamma_f,g_f)$ recovers to $[0,\pi/2]$, which is the same as the noncritical quench,
so the evolving Bloch vector $\hat{d}^t$ can sweep the whole Bloch sphere, thus we get $C_{\rm dyn}=1$.

%(1,-1,) to (1,1) ?

{\it Summary and discussion.}---In summary, we have studied the critical quantum quench, 
in which the prequenched Hamiltonian, or the postquenched Hamiltonian,
or both of them are set to be the critical points of equilibrium quantum phase transitions.
We demonstrate that the singularity of the Bogoliubov angle at the gap-closing momentum $k_c$ may lead to interesting topological properties
in such type of quantum quench.
The singularity of this momentum can lead to discontinuous change of the Pancharatnam geometrical phase of Loschomidt amplitude, 
which is the origination of the unquantization of DTOP;
if the momentum happens to be one of the fixed point of the Pancharatnam phase, then the DTOP can be half-quantized.
For the dynamical Chern number, the trajectories of the evolving Bloch vector $\hat{d}^t$ for $k\rightarrow k_{c-}$ and $k\rightarrow k_{c+}$ cut open 
the Bloch sphere, so the mapping from the $(k,t)$ space to the surface of the Bloch sphere is incomplete, therefore the dynamical Chern number may be 
 half-quantized or unquantized. The coincidence of the gap-closing momentum and the fixed point is a necessary condition for half-quantized dynamical
Chern number, because in the limit of this momentum, the  trajectory of the evolving Bloch vector $\hat{d}^t$ is exactly a meridian,
 which cut off half of the sphere.

The existence of the DQPT is also closely related to the singularity of the Bogoliubov angle, from the example shown in Fig. \ref{figB},
we can see that if $k^*$ coincides with the gap-closing momentum $k_c$, 
it does not lead to any singularity in rate function, and the DQPT is absent,
however, the DQPT can be recovered if there exists another $k^*$ that does not coincide with $k_c$.
We can also see that the effecs of the singularity of the Bogoliubov angles can even be canceled out 
if both the prequenched Hamiltonian and postquenched Hamiltonian 
are critical and the two gap-closing momenta are the same, in this case, the DTOP and dynamical Chern number restore to integer ones.

Although DQPTs have been studied extensively\cite{extIsing, XYDQPT, kitaev, rand, creutz, sw,ulink,sc,
mixed1,mixed2,excited,nonint,open,floquet},
the quench from or to a critical point, although very simple, is almost not well studied.
Our study demonstrates that such type of quantum quench is very important and valuable of further study, especially in the 
aspect of topological properties.
The conclusions in this paper are applicable for the integrable system like the XY chain with chiral symmetry, 
it is straightforward to generalize the study to other chiral symmetric systems, 
such as some exactly solvable spin chains\cite{cising,cisingI} and free fermion models. 
If there is no chiral symmetry, the conclusion of this paper may not hold, but technically, 
it is easy to try and expect some more interesting results. 
For some models like the XXZ chain which can only be exactly solved by Bethe Ansatz, as well as the models that are nonintegrable,
the calculation of DTOP and dynamical Chern number is a very interesting and important problem 
for both critical quench and noncritical quench, although it is currently unknown how to solve it.
The critical quantum quench is also very possible to be realized in the experiments of ultra-cold-atomic gases\cite{cold1,cold2}
and trapped ions\cite{trap1,trap2,trap3}, because the noncritical quench has already been realized, 
what we have to do is to tune the parameters to get a critical point in the experiment.

Finally, we need to emphasize again that the DTOP defined by Eq. (\ref{vD}) is problematic for the critical quench from the perspective of topology,
because the singularity of the Bogoliubov angle can make the manifold of the integral not closed.
Although the simple strategy demonstrated by Eqs. (\ref{Rk}) and (\ref{vD1})
can redefine the DTOP on a closed manifold, it does not change the final conclusions, i.e., the DTOP is still unquantized in certain cases, 
which is not satisfactory. 
If one can define a DTOP on a closed manifold by a completely different way, 
it is very worthwhile to look forward to it; in particular, for the case of Fig. \ref{cncc}(a), 
we expect a new DTOP that can be invariant in the same time interval $[t_n, t_{n+1}]$.
In addition, we can see that the dynamical Chern number defined by Eq. (\ref{chern}) has the similar problem, 
because the discontinuous change of the Bogoliubov angle at the gap-closing momentum can make the mapping from the 
space $(k,t)$ to the surface of the Bloch sphere incomplete, i.e, the dynamical Chern number is also defined on a unclosed manifold;
therefore, our findings also call for a new definition of the dynamical Chern number.
However, these are not simple tasks, they are out of the scope of current paper.

%{Acknowledgment}
This work is supported by the National Science Foundation of China (NSFC) under Grant Numbers 11975024, 11774002 and 11804383 
and the Anhui Provincial Supporting Program for Excellent Young Talents in Colleges and Universities under Grant No. gxyqZD2019023.


\begin{thebibliography}{widest-label}
\bibitem{Heyl2013}
M. Heyl, A. Polkovnikov, and S. Kehrein, Phys. Rev. Lett. {\bf 110}, 135704 (2013).
\bibitem{univ}
M. Heyl, Phys. Rev. Lett. {\bf 115}, 140602 (2015).
\bibitem{univ1}
A. Khatun and S. M. Bhattacharjee, Phys. Rev. Lett. {\bf 123}, 160603 (2019).
\bibitem{univ2}
Yantao Wu, Phys. Rev. B {\bf 101}, 014305 (2020).
\bibitem{univ3}
Yantao Wu, Phys. Rev. B {\bf 101}, 064427 (2020).
\bibitem{extIsing}
S. Bhattacharjee and A. Dutta, Phys. Rev. B {\bf 97}, 134306 (2018).
\bibitem{XYDQPT}
S. Vajna and B. D\'{o}ra, Phys. Rev. B {\bf 89}, 161105(R) (2014).
\bibitem{kitaev}
M. Schmitt and S. Kehrein, Phys. Rev. B {\bf 92}, 075114 (2015).
\bibitem{rand}
J. C. Halimeh, N. Yegovtsev, and V. Gurarie, Dynamical, quantum phase transitions in many-body localized systems, arXiv:1903.03109 (2019).
\bibitem{creutz}
R. Jafari, Henrik Johannesson, A. Langari, and M. A. Martin-Delgado, Phys. Rev. B {\bf 99}, 054302 (2019).
%\bibitem{nonhermitian}
%Longwen Zhou, QingHai Wang, Hailong Wang, and Jiangbing Gong, Phys. Rev. A {\bf 98}, 022129 (2018).
\bibitem{mixed1}
M. Heyl and J. C. Budich, Phys. Rev. B {\bf 96}, 180304(R) (2017).
\bibitem{mixed2}
U. Bhattacharya, S. Bandyopadhyay, and A. Dutta, Phys. Rev. B {\bf 96}, 180303(R) (2017). 
\bibitem{open}
Haifeng Lang, Yixin Chen, Qiantan Hong, and Heng Fan, Phys. Rev. B {\bf 98}, 134310 (2018).
\bibitem{floquet}
Kai Yang, Longwen Zhou, Wenchao Ma, Xi Kong, Pengfei Wang, Xi Qin, Xing Rong, Ya Wang, Fazhan Shi, Jiangbin Gong, 
and Jiangfeng Du, Phys. Rev. B {\bf 100}, 085308 (2019).
\bibitem{sw}
T. V. Zache, N. Mueller, J. T. Schneider, F. Jendrzejewski, J. Berges, and P. Hauke, Phys. Rev. Lett. {\bf 122}, 050403 (2019).
\bibitem{ulink}
Yi-Ping Huang, D. Banerjee, and M. Heyl, Phys. Rev. Lett. {\bf 122}, 250401 (2019).
\bibitem{sc}
A. Deshpande, B. Fefferman, M. C. Tran, M. Foss-Feig, and A. V. Gorshkov, Phys. Rev. Lett. {\bf 121}, 030501 (2018).
%\bibitem{inte}
%P. Jurcevic, H. Shen, P. Hauke, C. Maier, T. Brydges, C. Hempel, B. P. Lanyon, M. Heyl, R. Blatt, and C. F. Roos, Phys. Rev. Lett. {\bf 119}, 080501 (2017).
\bibitem{excited}
T. Tian, H.-X. Yang, L.-Y. Qiu, H.-Y. Liang, Y.-B. Yang, Y. Xu, and L.-M. Duan, Phys. Rev. Lett. {\bf 124}, 043001 (2020).
\bibitem{nonint}
I. Hagym\'{a}si, C. Hubig, \"{O}. Legeza, and U. Schollw\"{o}ck, Phys. Rev. Lett. {\bf 122}, 250601 (2019).
\bibitem{cold1}
M. Greiner, O. Mandel, T W H"{a}nsch, and I Bloch,  Nature {\bf 419}, 51 (2002).
\bibitem{cold2}
M. Cheneau, P. Barmettler, D. Poletti, M. Endres, P. Schauss, T. Fukuhara, C. Gross, I Bloch, C. Kollath, and S. Kuhr, Nature {\bf 481}, 484 (2012).
\bibitem{trap1}
P. Jurcevic, H. Shen, P. Hauke, C. Maier, T. Brydges, C. Hempel, B. P. Lanyon, M. Heyl, R. Blatt, and C. F. Roos, 
Phys. Rev. Lett. {\bf 119}, 080501 (2017).
\bibitem{trap2}
J. W. Britton, B. C. Sawyer, A. C. Keith, C. C. J. Wang, J. K. Freericks, H. Uys, M. J. Biercuk, and J. J. Bollinger, Nature {\bf 484}, 489 (2012).
\bibitem{trap3}
P. Jurcevic, B. P. Lanyon, P. Hauke, C. Hempel, P. Zoller, R. Blatt, and C. F. Roos, Nature {\bf 511}, 202 (2014).
\bibitem{DTOP}
J. C. Budich and M. Heyl, Phys. Rev. B {\bf 93}, 085416 (2016).
\bibitem{DPchern}
Ce. Wang, Pengfei Zhang, Xin Chen, Jinlong Yu, and Hui Zhai, Phys. Rev. Lett. {\bf 118}, 185701 (2017).
\bibitem{DPchern1}
C. Yang, Linhui Li, and Shu Chen, Phys. Rev. B {\bf 97}, 060304(R) (2018).
\bibitem{fidelity}
Wen-Long You, Ying-Wai Li, and Shi-Jian Gu, Phys. Rev. E {\bf 76}, 022101 (2007).
\bibitem{lkitaev}
O. Viyuela, D. Vodola, G. Pupillo, and M. A. Martin-Delgado1, Phys. Rev. B {\bf 94}, 125121 (2016).
\bibitem{cising}
S. M. Giampaolo and B. C. Hiesmayr, Phys. Rev. A {\bf 92}, 012306 (2015).
\bibitem{cisingI}
Chengxiang Ding, Phys. Rev. E {\bf 100}, 042131 (2019).
\end{thebibliography}
\end{document}